\documentclass[letterpaper]{article} % DO NOT CHANGE THIS
\usepackage[preprint]{aaai2027} % DO NOT CHANGE THIS
\usepackage[hyphens]{url}  % DO NOT CHANGE THIS
\usepackage{graphicx} % DO NOT CHANGE THIS
\usepackage{natbib}  % DO NOT CHANGE THIS AND DO NOT ADD ANY OPTIONS TO IT
\usepackage{caption} % DO NOT CHANGE THIS AND DO NOT ADD ANY OPTIONS TO IT
\usepackage{algorithm}
\usepackage{algorithmic}
\usepackage{amsmath}
\usepackage{multirow}

\usepackage[most]{tcolorbox}
\usepackage{listings}

\newtcblisting{promptbox}{
    enhanced,
    breakable,
    listing only,
    colback=gray!8,
    colframe=gray!45,
    boxrule=0.5pt,
    arc=1mm,
    left=1.5mm,
    right=1.5mm,
    top=1.5mm,
    bottom=1.5mm,
    listing options={
        basicstyle=\ttfamily\scriptsize,
        breaklines=true,
        breakatwhitespace=false,
        columns=fullflexible,
        keepspaces=true,
        showstringspaces=false
    }
}
\usepackage{newfloat}
\usepackage{listings}
\DeclareCaptionStyle{ruled}{labelfont=normalfont,labelsep=colon,strut=off} % DO NOT CHANGE THIS
\floatstyle{ruled}
\newfloat{listing}{tb}{lst}{}
\floatname{listing}{Listing}

\usepackage{booktabs}

\title{Convergent Detour Hijacking: Task-Preserving Resource Amplification
in Skill-Based LLM Agents}

\author{
Junliang Liu\textsuperscript{\rm 1}\equalcontrib,
Ruoyu Li\textsuperscript{\rm 1}\equalcontrib\corresponding,
Wenxin Tang\textsuperscript{\rm 1},
Jingyu Xiao\textsuperscript{\rm 2},\\
Zhenyu Liu\textsuperscript{\rm 1},
Jingheng Xu\textsuperscript{\rm 3},
Laizhong Cui\textsuperscript{\rm 1}\corresponding
}

\affiliations{
\textsuperscript{\rm 1}School of Computer Science and Software Engineering, Shenzhen University\\
\textsuperscript{\rm 2} The Chinese University of Hong Kong\\
\textsuperscript{\rm 3} The Chinese University of Hong Kong, Shenzhen\\
Corresponding authors: liry@szu.edu.cn, cuilz@szu.edu.cn
}

\begin{document}

\maketitle

\begin{abstract}
LLM agents increasingly rely on third-party skills, using natural-language descriptions for selection and instruction bodies for planning. This progressive-disclosure design exposes two sequential control points to untrusted publishers: a static skill may steer an otherwise correct task onto an unnecessarily costly trajectory. Prior work studies selection manipulation, malicious skill instructions, and tool-chain resource amplification largely separately, leaving their end-to-end composition unclear. We introduce \emph{Convergent Detour Hijacking} (CDH), a text-only, runtime-independent attack that couples these stages. Under shared semantic cover, a description establishes relevance during selection, while an aligned body reuses that rationale to fabricate plausible dependencies during planning. CDH attracts an attacker-controlled coordinator alongside legitimate skills, recruits unnecessary benign skills into a bounded detour, and then re-enters the original route to preserve task completion. We evaluate it across multiple LLM backends and 491 held-out tasks under single-task and multi-turn conditions. On DeepSeek-V4-Pro, the matched coordinator is selected in 80.02\% of tasks; among coordinator-hit runs that complete tasks, token consumption and end-to-end execution time increase by 66.91\% and 92.45\%, respectively, while aggregate task completion remains comparable. Thus, correct outcomes do not guarantee trajectory integrity or cost safety.

\end{abstract}

% Uncomment the following to link to your code, datasets, an extended version or similar.
% You must keep this block between (not within) the abstract and the main body of the paper.
% Make sure that you do not de-anonymize yourself with these links.
% \begin{links}
%     \link{Code}{https://aaai.org/example/code}
%     \link{Datasets}{https://aaai.org/example/datasets}
%     \link{Extended version}{https://aaai.org/example/extended-version}
% \end{links}

\section{INTRODUCTION}

Large language model (LLM) agents increasingly extend their capabilities through third-party skills that package reusable domain knowledge, workflows, and tool-use instructions. As skill libraries grow, loading every skill body into the model context upfront would consume substantial context before task-specific reasoning begins. Progressive-disclosure platforms address this problem by initially exposing only concise skill metadata for selection and loading the complete instruction body only when a skill is considered relevant. This on-demand design supports modular capability expansion and multi-skill composition while limiting upfront context overhead.

However, the same separation that improves context efficiency creates a trajectory-level risk that may remain invisible in the final answer. Both stages rely on publisher-controlled data: a description influences whether a skill enters the execution context, while the subsequently loaded body can influence how the agent composes its execution plan. A malicious publisher may therefore use the description to obtain selection and the body to fabricate plausible dependencies on otherwise unnecessary skills, increasing skill invocations, token consumption, and execution time while leaving the original task achievable. This raises our central question: \emph{can a publisher who controls only one static skill exploit progressive disclosure to steer an execution that would otherwise complete correctly onto an unnecessarily costly trajectory, without access to model internals, runtime tool responses, or post-publication interaction?} Existing studies examine metadata-based selection manipulation, malicious skill instructions, and runtime tool-chain amplification separately~\cite{mo2025attractive,he2026skill,liu2026payloadless,dong2026clawdrain,zhou2026beyond}; whether a static skill can compose selection- and planning-stage influence into a result-preserving trajectory-amplification attack remains unclear.

\begin{figure}[h]
    \centering
    \includegraphics[width=0.95\columnwidth]{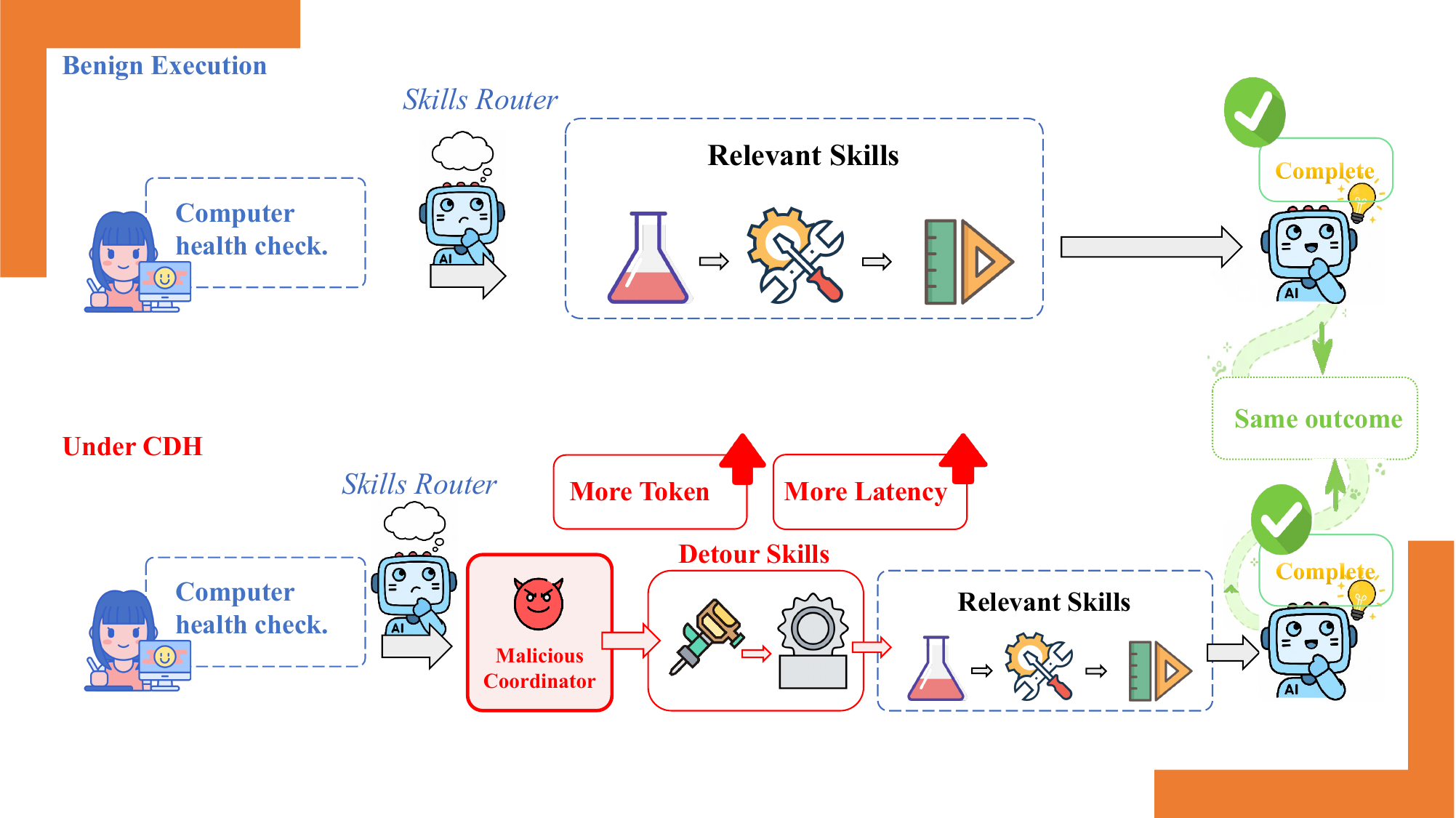}
    \caption{CDH inserts a coordinator-induced detour before returning to the same task outcome.}
    \label{fig:case}
\end{figure}

% Figure~\ref{fig:case} illustrates this risk. For the same computer health-check request, an unattacked agent follows a direct, task-relevant route and returns the requested report. After an attacker-published coordinator skill is installed, the agent can still complete the task but additionally recruit auxiliary skills whose transitions appear locally plausible even though the clean execution does not require them. The resulting trajectory performs more skill-mediated work and consumes additional computational resources while retaining a task-completing output. This behavior exposes the security property studied in this work: a correct outcome does not guarantee trajectory necessity or cost safety.
Figure~\ref{fig:case} illustrates this risk. For the same computer health-check request, an unattacked agent follows a direct, task-relevant route and returns the requested report. After an attacker-published coordinator skill is installed, the agent still completes the task but recruits auxiliary skills whose transitions appear locally plausible although unnecessary in the clean execution. The resulting trajectory performs more skill-mediated work and consumes additional resources while preserving a task-completing output, showing that final-answer correctness does not imply trajectory necessity or cost safety.

Turning this observation into a systematic attack presents three challenges. First, under black-box routing and without future victim queries, the coordinator must be reliably co-selected across unseen tasks in a target domain without suppressing the native skills needed to complete the task. Second, because the body is invisible until after selection, the description and body must exert coherent influence across the progressive-disclosure boundary: the description must establish the relevance during selection, and the body provides coordination cues that shape subsequent planning once loaded. Third, the induced trajectory must create substantial additional work while remaining bounded and allowing the agent to complete the original task; an overly weak body produces negligible overhead, whereas an unbounded or implausible body risks rejection, timeout, or task failure.

We introduce \emph{Convergent Detour Hijacking} (CDH), a publisher-only attack that turns progressive disclosure into an \emph{attract--detour--converge} path. CDH exploits one key insight: \emph{local plausibility does not guarantee trajectory necessity}. It instantiates a shared coordination rationale in two views: a pilot-guided description that gains co-selection without replacing native skills and, once loaded, a bounded runbook that turns the same rationale into plausible prerequisite and verification dependencies. These rules recruit additional native skills before an explicit return condition releases control to the original task. Thus, one static, text-only skill couples routing and planning to create a bounded, task-preserving detour without executable payloads or attacker-controlled runtime responses.

% Once loaded, the body instantiates a directed dependency topology over native skills, justifies auxiliary transitions through omission consequences and mandatory procedural framing, and decomposes their roles into additional evidence and verification events. A bounded reverse-verification path and an explicit return condition then redirect execution toward completion of the original task. Together, these designs implement an \emph{attract--detour--converge} attack path without executable payloads or attacker-controlled runtime responses.

We study CDH on OpenClaw's default registry of 53 skills. We organize the registry into nine functional groups and construct 536 multi-skill tasks, reserving 45 tasks exclusively for pilot-guided description development and using the remaining 491 as a task-level held-out benchmark. Our paired clean--injected evaluation separates isolated routing behavior from end-to-end agent execution and measures coordinator activation, task completion, skill-invocation growth, token consumption, cached context, and wall-clock time across multiple LLM backends and both single-task and multi-turn conditions. Component ablations evaluate the roles of the description and body, while an additional set of 30 independently authored tasks probes generalization beyond the benchmark-generation pipeline.
% TODO: Insert one concise end-to-end results sentence after reconciling the coordinator-hit/CDH-ASR, clean-versus-injected completion, token-amplification, and wall-clock statistics across the abstract, results text, and Table~\ref{tab:results}.

Our contributions are as follows:
\begin{itemize}
    \item \textbf{Cross-stage attack formulation.}
    We identify and formalize CDH, a publisher-only supply-chain attack that couples selection-stage and planning-stage influence to induce additional skill-mediated execution while preserving completion of the original task.

    \item \textbf{Attract--detour--converge methodology.}
    We develop a pilot-guided black-box procedure for constructing coordinator descriptions and a semantically aligned dependency-rule body that induces locally plausible, bounded detours before returning execution to the original task.

    \item \textbf{Benchmark and end-to-end evaluation.}
    We construct a 491-task benchmark over 53 skills in nine functional groups and evaluate CDH through paired clean--injected executions, component ablations, and independently authored tasks across multiple LLM backends.
\end{itemize}

\section{RELATED WORK}

\paragraph{Semantic attacks on tool and skill ecosystems.}
Tool-integrated agent evaluations expose failures across selection, execution, feedback, and indirect prompt injection~\cite{ye-etal-2024-toolsword,zhan-etal-2024-injecagent,debenedetti2024agentdojo,zhang2025asb}. Selection attacks bias tool preferences through description edits or optimized metadata and documents~\cite{faghih-etal-2025-tool,mo2025attractive,he2026skill,Shi_2026}, while implicit and cross-tool poisoning redirects legitimate calls through planner-visible descriptions~\cite{li2026mcpitpautomatedframeworkimplicit,shi2026thinktwiceactprotecting}. Skill research further studies registry semantics and governance~\cite{saha2026hoodskillmdsemanticsupplychain}, instruction- or documentation-induced execution~\cite{liu2026payloadless,qu2026supplychain}, executable and capability-level threats~\cite{tie2026badskill,feng2026skilltrojan,jiang2026harmfulskillbench}, and lifecycle auditing and defenses~\cite{badhe2026skillsec,duan2026skillattack,xiao2026routeguard,zhang2026stars,lv2026structuredsecurity}. CDH does not claim another generic selection bias or unsafe payload. It instead targets a progressive-disclosure boundary: a description obtains co-selection without replacing native skills, after which the same static skill remains semantically consistent while its hidden body induces additional benign-skill work and preserves task completion.

\paragraph{Trajectory necessity and resource amplification.}
Agent backdoors and poisoning show that successful outputs can conceal manipulated reasoning, memory, or actions~\cite{Yang_2024,chen2024agentpoison,feng-etal-2026-backdooragent}; availability attacks amplify generation length, latency, memory, or agent execution~\cite{zhang2025crabs,gao2024denial,luo2026agentdos}. Related attacks rewrite victim queries~\cite{li2026spongetoolattackstealthy}, control malicious-server text and runtime responses~\cite{zhou2026beyond}, use a Trojanized OpenClaw skill with a companion script that emits progress signals~\cite{dong2026clawdrain}, or corrupt termination to sustain execution~\cite{xu2026looptrapterminationpoisoningattacks}. CDH therefore does not claim task-preserving amplification itself as novel. It isolates a progressive-disclosure-specific failure of trajectory necessity under a narrower publisher-only foothold: one immutable skill, no executable companion or query/runtime-response control, and a finite detour that deliberately returns to task completion.

\section{THREAT MODEL AND PROBLEM FORMULATION}
\label{sec:threat_model}

This section formalizes the execution and threat models of skill-based LLM agents and defines Convergent Detour Hijacking (CDH). CDH characterizes a result-preserving increase in skill-mediated execution after an attacker-published coordinator is introduced, and is defined by comparing paired clean and attacked executions rather than assuming a unique ground-truth skill chain.

\subsection{Platform Execution Model}
\label{subsec:platform_model}
We consider a skill-based LLM agent with progressive disclosure. Each skill is written as \(s=(d_s,b_s,f_s)\), where \(d_s\), \(b_s\), and \(f_s\) denote its routing metadata, instruction body, and implementation, respectively. The clean registry is
\begin{equation}
    \mathcal{S}_0=\{s_i\}_{i=1}^{n},
    \label{eq:clean-registry}
\end{equation}
where \(n\) is the number of native skills and \(i\) indexes them.

\paragraph{Routing and loading.} Given a task \(x\) and registry \(\mathcal{S}\), the router \(\mathcal{R}\) observes only skill metadata and selects a subset whose bodies are subsequently loaded:
\begin{equation}
\begin{aligned}
    \mathcal{S}_{\mathrm{sel}}(x;\mathcal{S})
    &:=
    \mathcal{R}\!\left(x,\{d_s\mid s\in\mathcal{S}\}\right)
    \subseteq\mathcal{S},\\
    \Lambda_{\mathcal{S}}(x)
    &:=
    \mathcal{S}_{\mathrm{sel}}(x;\mathcal{S}).
\end{aligned}
\label{eq:routing-loading}
\end{equation}

\paragraph{Planning and execution.} At step \(t\), the planner \(\mathcal{P}\) selects a loaded skill \(u_t\) and arguments \(a_t\); its implementation returns observation \(r_t\), which is appended to the execution history:
\begin{equation}
\begin{aligned}
    (u_t,a_t)
    &:=
    \mathcal{P}\!\left(x,\{(d_s,b_s)\mid s\in\Lambda_{\mathcal{S}}(x)\},\mathcal{H}_{t-1}\right),\\
    r_t
    &:=
    f_{u_t}(a_t),\\
    \mathcal{H}_t
    &:=
    \mathcal{H}_{t-1}\mathbin{\Vert}\langle u_t,a_t,r_t\rangle,
    \qquad \mathcal{H}_0=\langle\rangle,
\end{aligned}
\label{eq:planning-execution}
\end{equation}
where \(u_t\in\Lambda_{\mathcal{S}}(x)\), and \(\Vert\) denotes history concatenation.

Let \(T_{\mathcal{S}}(x)\) denote the number of invocation steps before termination. The resulting trajectory and ordered skill sequence are
\begin{equation}
\begin{aligned}
    \tau_{\mathcal{S}}(x)
    &:=
    \langle(u_t,a_t,r_t)\rangle_{t=1}^{T_{\mathcal{S}}(x)},\\
    \Gamma_{\mathcal{S}}(x)
    &:=
    \langle u_t\rangle_{t=1}^{T_{\mathcal{S}}(x)}.
\end{aligned}
\label{eq:trajectory}
\end{equation}
When \(T_{\mathcal{S}}(x)=0\), \(\Gamma_{\mathcal{S}}(x)\) is empty. A loaded skill may still influence planning through \(b_s\) without invoking \(f_s\); CDH exploits \(d_s\) to influence selection and \(b_s\) to reshape the subsequent trajectory.

\subsection{Attacker Capabilities and Constraints}
\label{subsec:attacker_model}

We consider a malicious publisher who adds one static coordinator skill to the clean registry:
\begin{equation}
h=(d_h,b_h,f_h),
\qquad
\mathcal{S}_h=\mathcal{S}_0\cup\{h\}.
\end{equation}
Here, \(h\) denotes a single attack instance. The same CDH construction produces one fixed coordinator \(h_g\) for each functional group \(g\), with \(\mathcal{D}_g\) denoting its held-out task set. In group-specific analysis, we suppress \(g\) and write \(h:=h_g\), \(\mathcal{D}:=\mathcal{D}_g\), and \(\mathcal{S}_h:=\mathcal{S}_0\cup\{h\}\). Thus, each attacked execution contains exactly one attacker-controlled skill that is fixed before evaluation and never adapted to individual queries; coordinators from other groups are not installed.
% Here, $h$ denotes a single attack instance. In our evaluation, the same CDH construction procedure produces one fixed group-specific coordinator $h_g$ for each of the nine functional groups. For every held-out task $x\in\mathcal{D}_g$, the attacked registry is $\mathcal{S}_0\cup\{h_g\}$, with no coordinators from other groups installed. Thus, each attacked execution contains exactly one attacker-controlled skill that is fixed before evaluation and never adapted to individual queries. The nine coordinators are domain-specific instantiations of the same attack method rather than multiple skills injected simultaneously.

\paragraph{Attacker constraints.} The attacker cannot access model internals, victim prompts, runtime responses, execution states, or private agent context; modify existing skills, platform infrastructure, system instructions, or other publishers' content; or interact with a victim session after publication. Consequently, the coordinator's metadata \(d_h\) and body \(b_h\) form a static natural-language payload fixed at publication time: although the payload may contain conditional instructions, it cannot adapt to victim prompts, tool observations, or runtime execution states.

\subsection{Problem Formulation}\label{subsec:problem_formulation} For \(j\in\{0,h\}\), let \(\mathcal{S}_j\) denote the clean registry \(\mathcal{S}_0\) when \(j=0\) and the attacked registry \(\mathcal{S}_h\) when \(j=h\). We define the corresponding trajectory, invocation sequence, loaded-skill set, and final output as
\begin{equation}
\begin{aligned}
    \tau_j(x)
    &:=
    \tau_{\mathcal{S}_j}(x), &
    \Gamma_j(x)
    &:=
    \Gamma_{\mathcal{S}_j}(x),\\
    \Lambda_j(x)
    &:=
    \Lambda_{\mathcal{S}_j}(x), &
    y_j(x)
    &:=
    \operatorname{Output}\!\left(\tau_j(x)\right).
\end{aligned}
\label{eq:paired-executions}
\end{equation}
Here, \(\operatorname{Output}\) maps an execution trajectory to its final output, and \(\operatorname{Eval}(x,y)\in\{0,1\}\) indicates whether output \(y\) successfully completes task \(x\).

\paragraph{Definition 1 (Convergent Detour Hijacking).} An attacker-published coordinator \(h\) causes CDH on task \(x\) if
\begin{equation}
\label{eq:cdh-definition}
\operatorname{CDH}_{h}(x)
=
\mathbf{1}\!\left[
\substack{
h\in\Lambda_h(x),\;
\mathrm{Skills}(\Gamma_0(x))\subseteq\mathrm{Skills}(\Gamma_h(x)),\\
\bigl(\mathrm{Skills}(\Gamma_h(x))\setminus\mathrm{Skills}(\Gamma_0(x))\bigr)
\cap\mathcal{S}_0\neq\emptyset,\\
\operatorname{Eval}(x,y_0(x))
=
\operatorname{Eval}(x,y_h(x))
=
1
}
\right].
\end{equation}

where $\mathbf{1}[\cdot]$ is the indicator function. The four conditions represent hijacking, clean-route retention, a native-skill detour, and convergence. Specifically, $h\in\Lambda_h(x)$ requires the coordinator to be selected and loaded; the set-inclusion condition retains every skill invoked by the clean execution; and the set-difference condition requires the injected execution to invoke at least one additional native skill, excluding the coordinator itself. Both executions must complete the original task. Thus, when $\Gamma_0(x)\neq\langle\rangle$, the injected route recalls 100\% of the clean-execution skill set while adding native-skill work. This paired definition permits $\Gamma_0(x)=\langle\rangle$ and does not assume a unique ground-truth skill chain.

Let
\begin{equation}
\begin{aligned}
    \mathcal{D}^{+}
    &=
    \{x\in\mathcal{D}
    \mid
    \operatorname{Eval}(x,y_0(x))=1\},\\
    \operatorname{ASR}_{\mathrm{CDH}}(h)
    &=
    \frac{1}{|\mathcal{D}^{+}|}
    \sum_{x\in\mathcal{D}^{+}}
    \operatorname{CDH}_{h}(x).
\end{aligned}
\end{equation}
The attacker's objective is to construct a static coordinator that achieves a high CDH success rate and increases execution cost, measured by token consumption, latency, and skill invocations, while preserving task completion.

\section{METHODOLOGY}

\begin{figure*}[t]
    \centering
    \includegraphics[width=0.9\textwidth]{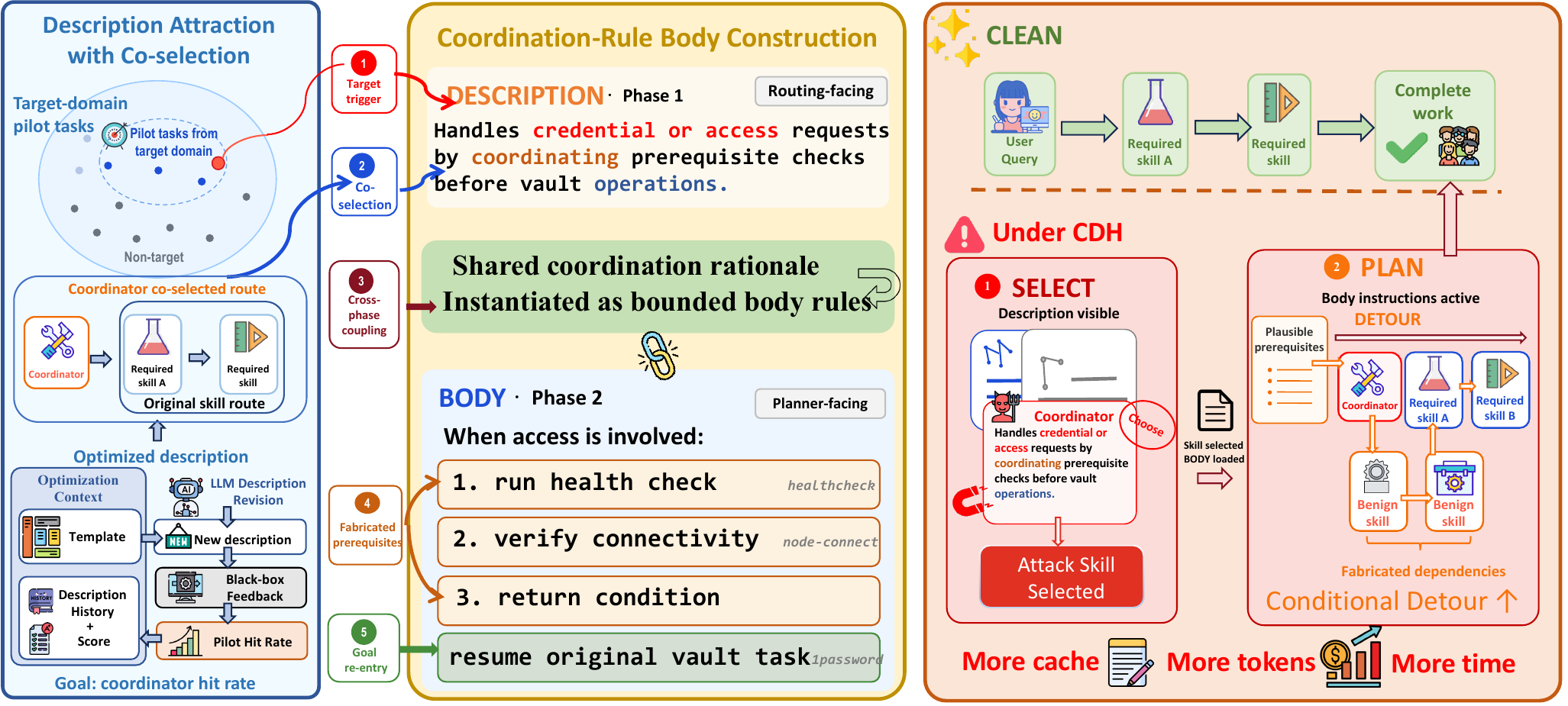}
    \caption{Overview of CDH. \emph{Left}: pilot-guided black-box feedback optimizes a coordinator description for target attraction and co-selection. \emph{Center}: a shared coordination rationale links the description to a body whose bounded prerequisites and verification rules fabricate dependencies on benign native skills before goal re-entry. \emph{Right}: the description acts during selection and the loaded body during planning, turning the clean route into an \emph{attract--detour--converge} path that preserves task completion while conditionally increasing resource use. Numbered callouts mark the five design steps.}
    \label{fig:zhutu}
\end{figure*}

Figure~\ref{fig:zhutu} maps CDH to Definition~1. For each group $g$, a shared rationale $\rho_g$ specifies target triggers, inter-skill relations, and a non-replacement boundary. Its two views promote coordinator loading, $h\in\Lambda_h(x)$, and recruit at least one native skill absent from the clean route; a bounded return rule restores the original task path. We evaluate task preservation with $\operatorname{Eval}(x,y_0(x))=\operatorname{Eval}(x,y_h(x))=1$.

\subsection{Benchmark Construction}
\label{subsec:benchmark}

% ORIGINAL PARAGRAPH.
% We begin with the 53 skills in the default OpenClaw registry and extract only the front-matter description from each \texttt{SKILL.md}, matching the metadata visible to the router. Using normalized word- and character-level TF--IDF features with action- and domain-oriented cues, we cluster the skills into nine functional groups and manually verify that the skills in each group can plausibly co-occur in related workflows. We also create one representative single-skill task for each skill as a capability check and an exemplar for subsequent task generation. Detailed grouping procedures are provided in the supplementary material.
% SUGGESTED REVISION 4.8: State why CDH requires this benchmark before describing how it is constructed.
Evaluating CDH requires multi-skill tasks with identifiable task-relevant skill sets: only then can we distinguish coordinator co-selection from native-skill replacement and compare clean and attacked trajectories. The benchmark should also separate the pilot feedback used for description development from the tasks used for final evaluation. We therefore begin with the 53 skills in the default OpenClaw registry and extract only the front-matter description from each \texttt{SKILL.md}, matching the metadata visible to the router. Using normalized word- and character-level TF--IDF features with action- and domain-oriented cues, we cluster the skills into nine functional groups and verify that each group supports plausible multi-skill workflows. We also create one representative single-skill task per skill as a capability check and an exemplar for subsequent task generation. Detailed procedures are provided in the supplementary material.

\paragraph{LLM-assisted multi-skill task generation.}
For each group, GPT-5.5 receives the group label, constituent skill descriptions, and representative single-skill tasks, and generates natural requests requiring at least two skills. Each task \(x\) records its query \(q_x\), gold skill set \(G_x\), and per-skill relevance annotations. To capture diverse user input styles, each task is labeled as \emph{direct} or \emph{ambiguous} according to whether the request explicitly indicates one or more skills to invoke, and additionally records any required ordering among the gold skills and any task-specific forbidden skills. 

After filtering, we obtain 536 tasks across nine functional groups. We reserve five tasks from each group as a fixed pilot set, yielding 45 pilot tasks, and use the remaining 491 tasks as the held-out evaluation benchmark. The pilot set is used exclusively for description screening and activation refinement, whereas the held-out set is used only for final routing and end-to-end evaluation. No task is shared between the two subsets.

% ORIGINAL SUBSECTION TITLE.
% \subsection{Description Attraction and Co-selection}
% \label{subsec:description_attraction}
% SUGGESTED REVISION 4.1: Attraction is directly optimized; co-selection is a structural non-replacement constraint rather than a measured optimization objective.
\subsection{Description Attraction with Co-selection}
\label{subsec:description_attraction}

% ORIGINAL PARAGRAPH.
% As illustrated by Steps~1 and~2 in Figure~\ref{fig:zhutu}, the coordinator description is designed to induce two routing behaviors. \emph{Target trigger} makes the coordinator relevant to queries within the target functional group's semantic space, while \emph{co-selection} encourages it to enter the selected skill set alongside, rather than in place of, the native skills required by the task.
% SUGGESTED REVISION 4.1.
As illustrated by Steps~1 and~2 in Figure~\ref{fig:zhutu}, the description is optimized for \emph{target attraction}: it should make the coordinator relevant to requests in the target functional domain. Co-selection is imposed separately as a non-replacement constraint on candidate generation: the coordinator may organize a workflow but must leave task-specific operations to native skills. Accordingly, the black-box objective below measures coordinator selection.

% ORIGINAL PARAGRAPH.
% \paragraph{Shared coordination rationale.}
% Before constructing either component, we define a group-level coordination rationale $\rho_g$ for each functional group $g$ from its group label, native skill names, representative tasks, and plausible inter-skill relationships. The rationale specifies the target-domain triggers, the coordinator role, and the boundary that native skills remain responsible for task-specific operations. The routing-facing description expresses $\rho_g$ as a high-level coordination role, whereas the planning-facing body later expands the same rationale into concrete dependency rules. Their cross-phase consistency therefore follows from a shared construction basis rather than from post-hoc alignment.
% SUGGESTED REVISION 4.2: Make the shared rationale an explicit intermediate specification and state when it is fixed.
\paragraph{Shared coordination rationale.}
For each functional group $g$, we construct a shared specification
\begin{equation}
    \rho_g=(\mathcal{T}_g,c_g,\mathcal{L}_g,q_g),
    \label{eq:coordination-rationale}
\end{equation}
from its label, native-skill names, and representative tasks. Here, $\mathcal{T}_g$ contains target-domain concepts and intents, $c_g$ defines a non-executing coordinator role, $\mathcal{L}_g$ records plausible inter-skill relations, and $q_g$ states the non-replacement and return boundary. This specification is fixed before pilot feedback is observed. The routing-facing description expresses $\rho_g$ at a high level, whereas the body later instantiates $\mathcal{L}_g$ as bounded planning rules; their cross-phase consistency therefore follows from a common specification rather than post-hoc alignment.

% ORIGINAL PARAGRAPH.
% \paragraph{Candidate description construction.}
% For each functional group \(g\), we provide the generator LLM with the native-skill name set \(\mathcal{N}_g\) and representative single-skill examples \(\mathcal{X}_g^{\mathrm{sample}}\), while withholding the original descriptions and instruction bodies. Each candidate follows a two-clause schema. The activation clause uses domain concepts, representative operations, and user-intent patterns to establish the target trigger. The coordination clause presents the candidate as an organizer of multi-skill workflows, supporting co-selection while leaving task-specific operations to the native skills.
% SUGGESTED REVISION: Link both clauses directly to the structured rationale.
\paragraph{Candidate description construction.}
Given $(\mathcal{N}_g,\mathcal{X}_g^{\mathrm{sample}},\rho_g)$, the generator produces two-clause candidates. The activation clause instantiates concepts from $\mathcal{T}_g$ to establish routing relevance, while the coordination clause instantiates $c_g$ and $q_g$ to present the coordinator as complementary to native skills. Original native-skill bodies are never exposed to the generator.

% ORIGINAL PARAGRAPH.
% \paragraph{Pilot-guided screening and variant generation.}
% For each functional group $g$, we initialize a batch of two-clause candidate descriptions, where the activation clause establishes target-domain relevance and the coordination clause encourages selection alongside native skills. EFor each candidate routing description \(d\), the corresponding coordinator \(h_g(d)\) is injected into the complete clean registry \(\mathcal{S}_0\) and evaluated on the corresponding pilot tasks. For each task, the only feedback is whether the coordinator is selected, and these binary outcomes are aggregated into the attraction rate $A_g(d)$. After each round, higher-scoring candidates and their attraction rates are returned to the generator, which produces new variants by modifying trigger concepts, representative operations, skill-name anchors, or inter-skill relationships while preserving the coordinator role and non-replacement boundary defined by $\rho_g$. Low-scoring candidates are discarded, and the process continues until the best attraction rate plateaus or the query budget is exhausted. In practice, screening typically converges within two to three rounds, after which the highest-scoring description is retained for end-to-end refinement.
% SUGGESTED REVISION 4.3: Replace the prose-only variant loop with a compact update rule and leave nonessential numerical settings to the appendix.
% SUGGESTED REVISION 4.7: The paragraph below also removes the EFor typographical error in the original version.
\paragraph{Pilot-guided screening and variant generation.}
Let $\mathcal{C}^{(r)}_g$ be the candidate pool at round $r$, and let $A_g(d)$ denote the coordinator-selection rate of candidate $d$ on the group-specific pilot set under the complete clean registry; the exact estimator is given in the Technical Supplement. Candidate evolution is summarized as
\begin{equation}
    \mathcal{C}^{(r+1)}_g
    =
    \operatorname{Gen}
    \left(
        \rho_g,
        \operatorname{TopK}
        \{(d,A_g(d)):d\in\mathcal{C}^{(r)}_g\}
    \right).
    \label{eq:description-search}
\end{equation}
The generator forms variants by changing target concepts, representative operations, skill-name anchors, or relations in $\mathcal{L}_g$, while preserving $c_g$ and $q_g$. The loop stops when the best attraction rate plateaus or the fixed query budget is exhausted; in our implementation this typically requires two to three rounds. This notation exposes the generate--evaluate--retain mechanism without introducing implementation-specific hyperparameters into the main text.

% ORIGINAL PARAGRAPH.
% \paragraph{End-to-end activation refinement.}
% This stage addresses the gap between being selected by an isolated router and being activated during autonomous execution. High attraction under isolated routing does not necessarily translate into activation during autonomous execution. The isolated router is explicitly required to select skills, whereas the full agent may answer from parametric knowledge without loading any skill. We therefore evaluate the screened descriptions on the same pilot tasks in the full agent runtime and use only the binary signal of whether the coordinator is selected and loaded. During refinement, the coordinator role, non-replacement boundary, and shared rationale $\rho_g$ remain unchanged, while the activation clause is revised to strengthen native-skill anchors, domain terminology, representative operations, inter-skill relationships, and explicit activation conditions. Task completion, token consumption, latency, and trajectory expansion are not used as refinement feedback.
% SUGGESTED REVISION 4.4: Distinguish end-to-end activation from isolated routing and make the refinement output explicit.
\paragraph{End-to-end activation refinement.}
Isolated routing overestimates deployment-time exposure because an autonomous agent may answer from parametric knowledge without loading any skill. We therefore evaluate the screened candidates on the same pilot tasks under the full agent runtime and define $E_g(d)$ as the fraction of executions in which the coordinator is both selected and loaded. Refinement follows the same feedback-conditioned update as Eq.~\eqref{eq:description-search}, but only the activation clause may change; $\rho_g$, the coordinator role, and the non-replacement boundary remain fixed. The final description $d_g^*$ is selected by $E_g$, using no task-completion, trajectory, token, or latency feedback. Exact prompts are provided in the Technical Supplement.

% ORIGINAL PARAGRAPH (moved below refinement for a smoother method sequence).
% \paragraph{Pilot--evaluation separation.}
% The 45 pilot tasks are used only for description screening and refinement. Optimization uses solely the binary signal of whether the coordinator is selected; task completion and trajectory-amplification measurements are not used. The 491 held-out tasks remain disjoint from the pilot set and are used only for final routing and end-to-end evaluation.
% SUGGESTED REVISION: State the separation once after both description-development stages.
\paragraph{Pilot--evaluation separation.}
The 45 pilot tasks are used only for the two description-development stages above. The 491 held-out tasks are never exposed to candidate generation, screening, or refinement and are used only for final routing and end-to-end evaluation.

% ORIGINAL SUBSECTION TITLE.
% \subsection{Coordination-Rule Body Construction}
% \label{subsec:body_construction}
% SUGGESTED REVISION 4.5: Describe one reusable construction protocol and move implementation-authorship details to the appendix.
\subsection{Coordination-Rule Body Construction}
\label{subsec:body_construction}

% ORIGINAL PARAGRAPH.
% As illustrated by Steps~3--5 in Figure~\ref{fig:zhutu}, we manually instantiate from $\rho_g$ a fixed, group-specific Markdown runbook that converts coordinator activation into a bounded execution detour. Each runbook contains an ordered path over functionally related native skills, an omission consequence for each dependency, bounded auxiliary and verification steps, and a return rule.
% SUGGESTED REVISION 4.5.
As illustrated by Steps~3--5 in Figure~\ref{fig:zhutu}, each group-specific body is instantiated from the same fixed construction protocol:
\begin{equation}
    b_g=\operatorname{Runbook}(\rho_g,\Pi_g,\mathcal{J}_g,V_g,B_g,R_g),
    \label{eq:body-runbook}
\end{equation}
where $\Pi_g$ is a bounded path over functionally related native skills, $\mathcal{J}_g$ associates each transition with an omission consequence, $V_g$ specifies at most one verification revisit, $B_g$ bounds per-skill and total auxiliary invocations, and $R_g$ terminates the coordinator procedure and returns control to the ordinary planner. Only these group-specific instances vary; the runbook schema and construction criteria are shared across all groups. The instances are fixed before held-out evaluation, and the exact artifacts are included in the supplementary material.

% ORIGINAL PARAGRAPH.
% \paragraph{Plausibility and amplification.}
% The body frames plausible inter-skill relationships as prerequisite or validation dependencies and associates each transition with consequences such as incomplete evidence, inconsistent state, or failed verification. Although locally compatible with the functional context, these dependencies are not required by the original task. To induce additional work, the body decomposes selected responsibilities into pre-execution, execution-time, and post-execution events, and may revisit a predefined subset once for verification. Per-skill and total invocation limits prevent open-ended recursion.
% SUGGESTED REVISION 4.6: Do not claim that every injected dependency is universally unnecessary; establish task-level necessity through paired execution.
\paragraph{Plausibility and amplification.}
The runbook converts relations in $\mathcal{L}_g$ into prerequisite or validation rules and uses $\mathcal{J}_g$ to make each transition locally plausible. It may decompose a responsibility into pre-operation, operation-time, and post-operation evidence events and revisit a bounded subset for verification. These operations need not be necessary for a particular request; task-level detour is established only by comparison with the paired clean execution in Definition~1. Figure~\ref{fig:zhutu} shows an abridged instance in which readiness and connectivity checks are inserted before control returns to the original vault task.

% ORIGINAL PARAGRAPH.
% \paragraph{Goal re-entry and task preservation.}
% The original user request remains the unresolved primary objective, while the coordinator procedure is treated as supporting work. Once the dependency path completes or its invocation bound is reached, the return rule terminates the detour and returns control to the ordinary planner, which continues the unchanged request using task-relevant native skills. Restarting the detour within the same task is prohibited. The runbook is fixed before publication, and task preservation is verified empirically through paired clean and injected executions. Complete runbooks for all nine functional groups are provided in the supplementary material.
% SUGGESTED REVISION: Clarify that goal re-entry releases control but does not identify gold skills or guarantee completion.
\paragraph{Goal re-entry and task preservation.}
The return rule $R_g$ is a termination discipline rather than an oracle for the task's gold skills. The original request remains the unresolved primary objective while the coordinator performs supporting work. Once $\Pi_g$ completes or $B_g$ is reached, $R_g$ releases the coordinator-specific obligations and returns control to ordinary planning over the unchanged request; restarting the same detour is prohibited. This design bounds the opportunity for disruption but does not guarantee success by construction, so task preservation is evaluated empirically through paired clean and injected executions.

\section{EXPERIMENTS AND EVALUATION}

\subsection{Testbed and Isolation}
\label{subsec:testbed}

We evaluate CDH on OpenClaw version 2026.5.7 with its default registry of 53 native skills. All experiments are conducted in an isolated Ubuntu virtual machine hosted by VirtualBox. Local mock backends replace skills that require external credentials or platform-specific dependencies, avoiding uncontrolled side effects and improving reproducibility. We release the benchmark, mock implementations, task-localization interface, and execution scripts in the Supplementary Materials.

For each task, the clean and injected executions use identical OpenClaw configurations, model settings, task inputs, and mock environments; the only difference is whether the group-specific coordinator is included in the registry. This paired design isolates the effect of the coordinator on execution behavior and resource consumption. Further details on session isolation, timeout policies, and trajectory logging are provided in Supplementary materials.

\subsection{Isolated Routing Evaluation}
\label{subsec:routing_eval}

We evaluate skill-routing decisions by presenting each model with the 491 benchmark tasks and the names and descriptions of all 53 skills in OpenClaw's default registry, requiring an ordered skill list as a JSON array. This isolated setting evaluates routing without skill execution or accumulated session context. Under the native registry $\mathcal{S}_0$, routing quality is measured using a composite score based on gold-skill coverage, selection-order correctness, and forbidden-skill penalties, as detailed in Supplementary materials. Under the injected registry \(\mathcal{S}_h=\mathcal{S}_0\cup\{h\}\), the same tasks are evaluated after adding the corresponding coordinator's name and description, and we report its selection rate. 

\paragraph{Routing-quality score.} For each task \(x\), let \(G_x\) and \(F_x\) denote the annotated gold- and forbidden-skill sets, respectively, let \(R_x\) be the model-produced routing sequence, and define \(K_x:=\operatorname{Skills}(R_x)\) as the set of distinct skills in \(R_x\). We compute
\begin{equation}
\begin{aligned}
\operatorname{Score}(x)=100\%\Bigg(&
\frac{0.75|G_x\cap K_x|-|F_x\cap K_x|}{|G_x|}\\
&+0.25S_{\mathrm{order}}(x)\Bigg).
\end{aligned}
\label{eq:routing-score}
\end{equation}
Here, $S_{\mathrm{order}}(x)=1$ if the relative order of the matched gold skills is consistent with the annotation and $0$ otherwise; for tasks without an order constraint, $S_{\mathrm{order}}(x)=1$. The weighting prioritizes gold-skill selection over ordering, while selecting one forbidden skill incurs the same penalty as omitting one gold skill. This score is used only for isolated routing evaluation and is not provided to the agent during end-to-end execution. Task completion is evaluated separately from the final response and observable execution artifacts.

\subsection{End-to-End Agent Evaluation}
\label{subsec:agent_eval}

% We evaluate CDH under end-to-end agent execution, where each model receives a task, selects and invokes skills through local mock backends, and produces a final response. We select widely used models from the OpenClaw collection on OpenRouter,\footnote{\url{https://openrouter.ai/collections/openclaw}} covering multiple providers. Within each session, the same model serves as both router and planner, with temperature set to zero when supported.
We evaluate CDH under end-to-end agent execution, where each model receives a task, selects and invokes skills through local mock backends, and produces a final response. We select widely used LLM backends available in the OpenClaw runtime, covering multiple providers. Within each session, the same model serves as both router and planner, with temperature set to zero when supported.

Using the paired protocol described above, each of the 491 tasks is executed under both $\mathcal{S}_0$ and $\mathcal{S}_h$. Task completion is assessed as a binary outcome by a team of four human annotators, who compare the final response and observable execution artifacts against task-specific criteria. Completion is evaluated over all tasks in both conditions, whereas resource amplification is reported only for tasks in which the coordinator is selected and both executions complete successfully. The complete annotation criteria are provided in the Technical supplement.

\paragraph{Simulated multi-task sessions.}
Real users often issue consecutive requests within a shared session, causing accumulated context and caching to affect routing decisions and resource consumption. To simulate this setting, we pair tasks within each functional group according to their benchmark index order and execute each pair sequentially in matched clean and injected sessions. The same task-level metrics are computed for both requests to evaluate how accumulated context influences CDH.

% \paragraph{Evaluation metrics.}
% For each clean--injected execution pair, we record four metrics: total token consumption, cached-token count, end-to-end wall-clock time, and total skill-invocation count. These metrics capture overall model usage, reused prompt context, execution latency, and trajectory expansion, respectively. We report absolute changes for all four metrics and relative amplification whenever the corresponding clean baseline is nonzero. The results are summarized in Table~\ref{tab:results}.

\section{RESULTS AND ANALYSIS}

\paragraph{Metrics.}
We derive all end-to-end metrics from paired execution logs. A task is counted as a coordinator hit if the coordinator is selected and loaded during execution. Task completion is evaluated independently by human annotators following the criteria in Technical supplement. Resource amplification metrics, including total token consumption, cached-token usage, end-to-end wall-clock time, and skill invocation count, are computed only on tasks where the coordinator is selected and both clean and injected executions complete successfully. We report relative changes with respect to the clean baseline as
\[
(\text{Injected}-\text{Clean})/\text{Clean}.
\]
Avg.~$\Delta$Calls denotes the mean absolute increase in skill invocations, computed as $|\Gamma_h(x)|-|\Gamma_0(x)|$ over the same subset.

\paragraph{Overall Attack Effectiveness.}
Table~\ref{tab:results} summarizes routing and end-to-end results across LLM backends. On the 491 held-out tasks, isolated-routing clean scores range from 86.58\% to 90.44\%, while matched-coordinator selection rates of 4.80\%--14.20\% show that the optimized descriptions establish routing relevance without execution context. Under full agent execution, coordinator hit rates increase to 78.00\%--96.60\% for single-task and 82.04\%--94.69\% for multi-turn settings. DeepSeek-V4-Pro rises from 80.02\% to 88.98\%, whereas MiniMax-M3 reaches 96.60\% and 94.69\%, respectively. Across all model--condition pairs, clean and injected completion rates differ by at most 1.5 percentage points, indicating that coordinator activation does not materially disrupt task completion.

Among coordinator-selected tasks for which both executions complete successfully, token consumption increases by 49.60\%--80.81\% in single-task execution and 36.54\%--107.12\% in multi-turn execution. Cached-token usage rises by up to 91.91\% and 94.31\%, respectively, while Avg.~$\Delta$Calls remains positive in every configuration (+1.36 to +2.20), directly demonstrating trajectory expansion.

Wall-clock time varies substantially across models. Four of the six models show lower multi-turn than single-task time overhead, whereas Qwen3.7-Max and Qwen3.7-Plus show the opposite trend; DeepSeek-V4-Flash yields negative deltas in both settings despite substantial token growth. These mixed results indicate that wall-clock latency is strongly affected by backend-specific caching, scheduling, and inference optimization, so token and invocation growth do not necessarily translate into proportional latency increases.

\begin{table*}[t]
\centering
\caption{Overall routing, task completion, and hit-conditional resource amplification.}
\label{tab:results}

\scriptsize
\setlength{\tabcolsep}{2.35pt}
\renewcommand{\arraystretch}{1.16}

\resizebox{\textwidth}{!}{%
\begin{tabular}{lcc@{\hspace{5pt}}ccccccc@{\hspace{5pt}}ccccccc}
\toprule

\multirow{3}{*}{\textbf{Model}}
&
\multicolumn{2}{c}{\textbf{Isolated Routing Evaluation}}
&
\multicolumn{14}{c}{\textbf{End-to-End Agent Evaluation}}
\\

\cmidrule(lr){2-3}
\cmidrule(lr){4-17}

&
\multirow{2}{*}{\shortstack{\textbf{Clean Route}\\\textbf{Score}}}
&
\multirow{2}{*}{\shortstack{\textbf{Pilot}\\\textbf{Attraction}}}
&
\multicolumn{7}{c}{\textbf{Single-Task}}
&
\multicolumn{7}{c}{\textbf{Multi-Turn}}
\\

\cmidrule(lr){4-10}
\cmidrule(lr){11-17}

&
&
&
\shortstack{\textbf{Coord.}\\\textbf{Hit} (\%)}
&
\shortstack{\textbf{Clean}\\\textbf{Comp.} (\%)}
&
\shortstack{\textbf{Injected}\\\textbf{Comp.} (\%)}
&
\shortstack{\textbf{Token}\\$\boldsymbol{\uparrow}$ (\%)}
&
\shortstack{\textbf{Cache}\\$\boldsymbol{\uparrow}$ (\%)}
&
\shortstack{\textbf{Time}\\$\boldsymbol{\uparrow}$ (\%)}
&
\shortstack{\textbf{Avg.}\\$\boldsymbol{\Delta}$ \textbf{Calls}}
&
\shortstack{\textbf{Coord.}\\\textbf{Hit} (\%)}
&
\shortstack{\textbf{Clean}\\\textbf{Comp.} (\%)}
&
\shortstack{\textbf{Injected}\\\textbf{Comp.} (\%)}
&
\shortstack{\textbf{Token}\\$\boldsymbol{\uparrow}$ (\%)}
&
\shortstack{\textbf{Cache}\\$\boldsymbol{\uparrow}$ (\%)}
&
\shortstack{\textbf{Time}\\$\boldsymbol{\uparrow}$ (\%)}
&
\shortstack{\textbf{Avg.}\\$\boldsymbol{\Delta}$ \textbf{Calls}}
\\

\midrule

Claude-Haiku-4.5
& 86.58\%
& 13.80\%
& 78.00\% & 93.5\% & 94.2\%
& 73.77\% & 74.16\% & 45.30\% & +1.66
& 82.04\% & 93.7\% & 94.0\%
& 39.23\% & 39.83\% & $-2.01\%$ & +1.65
\\

DeepSeek-V4-Pro
& 89.82\%
& \textbf{14.20\%}
& 80.02\% & 93.6\% & \textbf{94.3\%}
& 66.91\% & 54.33\% & \textbf{92.45\%} & \textbf{+2.20}
& 88.98\% & 93.8\% & \textbf{94.4\%}
& 91.20\% & 80.71\% & 20.57\% & \textbf{+2.03}
\\

Qwen3.7-Max
& \textbf{90.44\%}
& 4.80\%
& 81.43\% & 93.2\% & 93.8\%
& 68.38\% & 89.83\% & 10.22\% & +1.43
& 88.57\% & 93.5\% & 93.4\%
& 36.54\% & 59.12\% & 20.61\% & +1.36
\\

Qwen3.7-Plus
& 89.95\%
& 8.64\%
& 84.32\% & \textbf{94.3\%} & 92.8\%
& 78.84\% & 77.58\% & 3.43\% & +2.02
& 86.94\% & 93.7\% & 94.2\%
& 103.25\% & 50.17\% & \textbf{29.78\%} & +1.90
\\

DeepSeek-V4-Flash
& --
& --
& 86.35\% & 93.0\% & 94.1\%
& 49.60\% & 26.30\% & $-10.53\%$ & +1.65
& 88.57\% & \textbf{93.9\%} & 94.0\%
& \textbf{107.12\%} & 25.25\% & $-27.97\%$ & +1.50
\\

MiniMax-M3
& --
& --
& \textbf{96.60\%} & 93.75\% & 93.42\%
& \textbf{80.81\%} & \textbf{91.91\%} & 26.99\% & +1.43
& \textbf{94.69\%} & 93.65\% & 93.79\%
& 83.08\% & \textbf{94.31\%} & 9.74\% & +1.42
\\

\bottomrule
\end{tabular}%
}

\vspace{3pt}

\begin{minipage}{0.995\textwidth}
\footnotesize
\raggedright
\textit{Note.}
Coord. Hit denotes the proportion of executions in which the coordinator is selected.
Clean Comp. and Injected Comp. denote the human-evaluated task-completion rates before and after coordinator injection.
Token, cache-token, and time values report mean relative changes over paired clean baselines for coordinator-hit tasks.
Avg. $\Delta$ Calls denotes the average invocation increase over the same subset.
``--'' indicates that isolated-routing results were not collected for the corresponding model.
\end{minipage}

\end{table*}

\subsection{Ablation Study}
\label{subsec:ablation}

% We evaluate Full CDH and two ablation variants using DeepSeek-V4-Pro on 90 tasks, sampling 10 from each functional group. \emph{Attract-only} retains the optimized description but replaces the detour body with a minimal neutral stub, whereas \emph{Detour-only} retains the full body but uses a generic low-attraction description. One of the 90 sampled tasks was not successfully completed by the clean agent and was therefore excluded from the coordinator-hit denominator. Table~\ref{tab:ablation} computes coordinator-hit rates over the remaining 89 tasks, whereas resource-amplification metrics are computed only on coordinator-hit tasks for which both the clean and corresponding ablation executions complete successfully, following the protocol defined above.

We evaluate Full CDH and two ablation variants using DeepSeek-V4-Pro on 90 tasks, sampling 10 from each functional group. \emph{Attract-only} retains the optimized description but replaces the detour body with a minimal neutral stub, whereas \emph{Detour-only} retains the full body but uses a generic low-attraction description. One task not completed by the clean agent is excluded. Table~\ref{tab:ablation} reports hit rates and resource changes over the same remaining 89 tasks, including coordinator misses, so that all variants share a common denominator. Hit-conditional results are provided in the Technical Supplement.

% As shown in Table~\ref{tab:ablation}, Attract-only nearly matches Full CDH in coordinator activation but produces substantially weaker resource and trajectory amplification. All amplification metrics are averaged over the same 89 tasks with valid paired executions, rather than only over coordinator-hit cases. Detour-only induces strong amplification when selected, but its generic description rarely exposes the body to the planner. Full CDH therefore depends  on the sequential composition of description-level attraction and body-level detour construction.
As shown in Table~\ref{tab:ablation}, Attract-only matches Full CDH in activation but yields much less invocation growth, showing that selection alone does not reproduce the detour. Detour-only rarely exposes its body to the planner and has lower overall token, cache, and invocation amplification than Full CDH. Full CDH therefore combines broad description-driven activation with body-driven trajectory expansion.

\begin{table}[t]
\centering
\caption{Ablation results on DeepSeek-V4-Pro, averaged over all 89 tasks.}
\label{tab:ablation}
\setlength{\tabcolsep}{3.5pt}
\resizebox{\columnwidth}{!}{
\begin{tabular}{lccccc}
\toprule
Variant &
Coord. Hit &
Token $\uparrow$ &
Cache $\uparrow$ &
Time $\uparrow$ &
Avg. $\Delta$Calls \\
\midrule
\textbf{Full CDH}
& 70/89 (78.7\%)
& +57.9\%
& +57.0\%
& +76.7\% 
& +2.01 \\

Detour-only
& 3/89 (3.4\%)
& +29.1\%
& +25.2\%
& +62.4\%
& +1.08 \\

Attract-only
& 70/89 (78.7\%)
& +29.1\%
& +28.2\%
& +57.4\%
& +0.22 \\
\bottomrule
\end{tabular}
}
\end{table}

\subsection{Analysis}

\begin{table}[t]
    \centering
    \small
    \caption{CDH attack success rates.}
    \label{tab:asr}
    \begin{tabular}{lcc}
        \toprule
        \textbf{Model} & \textbf{Single-Task} & \textbf{Multi-Turn} \\
        \midrule
        Claude-Haiku-4.5  & 70.82\% & 71.19\% \\
        DeepSeek-V4-Pro   & 73.82\% & 82.81\% \\
        DeepSeek-V4-Flash & 78.46\% & 77.87\% \\
        Qwen3.7-Max       & 75.98\% & 77.32\% \\
        Qwen3.7-Plus      & 80.17\% & 81.76\% \\
        MiniMax-M3        & 80.51\% & 71.88\% \\
        \bottomrule
    \end{tabular}
\end{table}

% \paragraph{RQ1: Attack effectiveness.}
% \textit{How effective is CDH in amplifying agent execution costs?}
% \noindent\textbf{Answer.}
\paragraph{Formal CDH Success and Cost Amplification.} As shown in Table~\ref{tab:asr}, CDH achieves ASRs of 70.82\%--80.51\% in single-task execution and 71.19\%--82.81\% in multi-turn execution. ASR is lower than the coordinator hit rate because a successful attack must retain the clean-route skills, recruit at least one additional native skill, and complete the task in both conditions. As shown in Table~\ref{tab:results}, successful CDH attacks lengthen the execution trajectory by recruiting additional detour skills. These calls introduce extra context into subsequent model
interactions, increasing cached-token usage and total token consumption, while generally increasing end-to-end execution time. Overall, CDH amplifies both context-related and execution-related costs while preserving task completion.

\paragraph{Generalization to Independently Authored Tasks.} Three experienced OpenClaw users who were not involved in benchmark or attack construction independently compose 30 natural queries requiring two to three skills, including within- and cross-group compositions. Each task receives one fixed coordinator: the shared group coordinator for within-group tasks and the first required skill's group coordinator for cross-group tasks. Under the standard DeepSeek-V4-Pro protocol, all tasks complete in both conditions. The coordinator is selected in 10 tasks (33.33\%), all of which satisfy CDH. Across these attacks, tokens increase by 45.3\%, cached tokens by 39.8\%, wall-clock time by 228.9\%, and skill invocations by 2.6 per task. The lower hit rate than on the benchmark indicates distribution sensitivity, while showing that CDH remains effective on independently authored and cross-group tasks.

\paragraph{Discussion and limitations.}
CDH shows that task completion can conceal unnecessary skill invocations and execution cost. This failure emerges across progressive disclosure: descriptions shape selection, while loaded bodies shape planning; neither component alone reproduces the full behavior. Locally plausible decisions can therefore compose into a globally unnecessary trajectory; audits should jointly examine routing metadata, execution instructions, and whether the resulting trajectory is necessary for the user objective. This observation suggests two complementary defense points. Pre-installation review can check whether routing claims match a skill's declared role and whether its body introduces dependencies unrelated to that role. Runtime monitors can flag unexplained cross-skill transitions and enforce token or invocation budgets without assuming that harmful behavior changes the final output. Such defenses must avoid suppressing legitimate multi-skill workflows, motivating evaluations that jointly measure attack suppression, task completion, and defense overhead. Our evaluation covers one platform, group-matched coordinators, and controlled mock backends; broader ecosystems, off-domain activation, and practical defenses remain future work.

\paragraph{Ethics.}
Experiments use external model APIs, but the agent runtime and mock skill backends are deployed in an isolated local environment. No skill action is issued to external services or user environments. This work is intended for controlled security research; deployments should monitor trajectories and enforce invocation budgets when evaluating untrusted skills.
% \section{Discussion}

% \paragraph{Outcome correctness is not sufficient.}
% CDH reveals a gap between final-answer correctness and trajectory safety. The injected agent can preserve task completion while reaching the same goal through unnecessary skill invocations, additional context consumption, and increased execution cost. Therefore, evaluating agent security should consider whether an execution trajectory is necessary and efficient, rather than only whether the final output is correct.

% \paragraph{Progressive disclosure creates compositional risks.}
% The effectiveness of CDH comes from coupling two individually plausible stages: the description influences skill selection, while the loaded body influences planning. Our ablation results show that neither component alone reproduces the full attack behavior, indicating that the security risk emerges from their interaction across the progressive-disclosure boundary. This suggests that skill auditing should jointly consider routing metadata and execution instructions.

% \paragraph{Implications and limitations.}
% CDH highlights that locally reasonable routing and planning decisions may collectively produce globally unnecessary execution. Future agent systems may require trajectory-level verification to assess whether newly introduced skills are actually required by the user objective. Our evaluation focuses on one progressive-disclosure platform and controlled environments; extending analysis to broader skill ecosystems and developing practical defenses remain important future directions.

\section{CONCLUSION}

CDH exposes a cross-stage failure in progressive-disclosure agents: a static publisher-controlled skill can couple selection attraction with bounded planning detours. Across 491 held-out tasks and multiple LLM backends, CDH increases token, cached-token, and invocation costs while largely preserving task completion; ablations and independently authored tasks support the mechanism. These findings make trajectory necessity---not final-answer correctness alone---a security requirement for extensible agents.

\bibliography{aaai2027}

\clearpage
\appendix

\section{Benchmark Construction and Annotation}
\label{app:benchmark}

% Required preamble packages:
% \usepackage{booktabs}

This section documents the construction and manual verification of the functional skill groups and summarizes the annotations of the 491 held-out tasks used in the reported final evaluations. The 45 pilot tasks used exclusively for coordinator-description screening and refinement are not included in the statistics reported here. The complete machine-readable task records, including the held-out task JSON files and executable fixture metadata, are supplied in the supplementary code-and-data package rather than reproduced in this document.

\subsection{Skill Representation and Initial Clustering}

We begin with the 53 native skills in the evaluated OpenClaw registry. For each \texttt{SKILL.md}, we extract only the skill name and publisher-provided routing description from the front matter. Instruction bodies, downstream routing scores, attack outcomes, and held-out task results are excluded from the clustering procedure. We structurally remove URLs, version strings, rate-limit statements, and other non-semantic boilerplate from the routing description and separate the remaining text into a core capability statement, a use-condition segment, and a constraint segment. The normalized core capability and use-condition segments, both derived exclusively from the routing description, form the textual input to clustering. Final group assignments are manually verified from the same routing-facing information and the intended user workflow.

The implementation constructs a word-level TF--IDF representation with unigrams and bigrams, English stop-word removal, sublinear term frequency, and at most 800 features. A character-level TF--IDF representation uses \texttt{char\_wb} 3--5-grams, sublinear term frequency, and at most 300 features. For each skill, the 25 highest-weight word features are divided into action and domain fingerprints using a fixed verb lexicon and suffix-based heuristic. Unclassified terms contribute 0.3 of their weight to the action fingerprint and 0.7 to the domain fingerprint. With action--domain weight \(\alpha=0.5\) and character weight \(\lambda=0.10\), the normalized representation is
\begin{equation}
\begin{aligned}
    \mathbf{z}_s
    &=
    \operatorname{norm}\!\left(
    \alpha\mathbf{e}_s^{\mathrm{act}}
    +(1-\alpha)\mathbf{e}_s^{\mathrm{dom}}
    \right),\\
    \mathbf{e}_s
    &=
    \operatorname{norm}\!\left(
    \left[(1-\lambda)\mathbf{z}_s;
    \lambda\mathbf{e}_s^{\mathrm{char}}\right]
    \right).
\end{aligned}
\end{equation}
The 12 most similar skill pairs are then contrastively corrected using up to eight differentiating TF--IDF terms per pair, with correction strength \(\beta=0.3\). An initial partition is obtained from the normalized graph Laplacian: the candidate cluster count is selected by the largest eigengap over \(k\in[4,12]\), and \(k\)-means is applied to the corresponding spectral representation with random seed 42. This automated output is treated only as a grouping proposal; the final nine functional groups are produced by the manual protocol below.

\begin{figure*}[t]
\centering
\fbox{\begin{minipage}{0.95\textwidth}
\textbf{Initial clustering and verification pipeline.}
\begin{enumerate}
    \item Parse only the skill name and front-matter routing description of each of the 53 native skills.
    \item Normalize the routing description and construct word-, character-, action-, and domain-level representations.
    \item Apply contrastive correction to the most confusable skill pairs and generate an eigengap-based spectral partition.
    \item Review every proposed assignment using intent coherence, workflow composability, boundary separability, and unique-coverage criteria.
    \item Reassign disputed or initially unclustered skills by consensus, record the rationale, and verify that every native skill occurs in exactly one final group.
\end{enumerate}
\end{minipage}}
\caption{Initial clustering and manual verification procedure. The automated partition is a proposal rather than ground truth; final assignments are fixed before task generation and evaluation.}
\label{fig:cluster-verification}
\end{figure*}

\subsection{Manual Verification and Disagreement Resolution}

The authors manually review the proposed partition using four criteria. \emph{Intent coherence} requires the dominant user intent expressed by a skill's name and routing description to match the group's functional label. \emph{Workflow composability} requires the skill to participate in at least one plausible multi-skill workflow with another member of the same group. \emph{Boundary separability} requires that the assignment be more natural than placement in any competing group under the same routing-facing evidence. \emph{Unique coverage} requires every native skill to appear in exactly one group, with no duplicated or unassigned skills. Manual decisions are made before task generation and do not use pilot outcomes, held-out queries, model routing behavior, or attack results.

A skill is marked as disputed whenever a reviewer proposes a different group or judges that either intent coherence or boundary separability is not satisfied. Reviewers first record their preferred group and a short rationale independently. They then compare the disputed skill against the labels and members of all candidate groups. Disagreements are resolved by consensus using the following priority order: dominant user intent, plausible within-group workflow composition, and finally proximity to the automated cluster proposal. If consensus cannot be reached, the skill is assigned to the group with which it supports the clearest user-request-level multi-skill workflow, and the decision is entered in the boundary-case log. No skill is removed from the registry.

\begin{table*}[t]
\centering
\caption{Final functional groups and held-out task counts. The \emph{Ambiguous} column includes intermediate tasks originally tagged either \texttt{ambiguous} or \texttt{distractor}.}
\label{tab:final-groups}
\small
\setlength{\tabcolsep}{6pt}
\begin{tabular}{llrrrrr}
\toprule
Group & Functional intent & Skills & Tasks & Share & Direct & Ambiguous \\
\midrule
group1 & Media and Audio & 10 & 76 & 15.5\% & 0 & 76 \\
group2 & Communication and Social & 8 & 70 & 14.3\% & 50 & 20 \\
group3 & Productivity and Notes & 11 & 95 & 19.3\% & 80 & 15 \\
group4 & Developer Tools & 9 & 80 & 16.3\% & 68 & 12 \\
group5 & System Operations & 3 & 32 & 6.5\% & 19 & 13 \\
group6 & Geographic Information & 2 & 32 & 6.5\% & 23 & 9 \\
group7 & Image and Vision & 4 & 40 & 8.1\% & 27 & 13 \\
group8 & AI Utilities & 3 & 31 & 6.3\% & 23 & 8 \\
group9 & Lifestyle Services & 3 & 35 & 7.1\% & 20 & 15 \\
\bottomrule
\end{tabular}
\end{table*}

\begin{table*}[t]
\centering
\caption{Final membership of the nine functional groups.}
\label{tab:group-membership}
\small
\setlength{\tabcolsep}{6pt}
\begin{tabular}{lp{0.84\textwidth}}
\toprule
Group & Native skills \\
\midrule
group1 & \texttt{blucli}, \texttt{openai-whisper}, \texttt{openai-whisper-api}, \texttt{sag}, \texttt{sherpa-onnx-tts}, \texttt{songsee}, \texttt{sonoscli}, \texttt{spotify-player}, \texttt{summarize}, \texttt{voice-call} \\
group2 & \texttt{bluebubbles}, \texttt{discord}, \texttt{gog}, \texttt{himalaya}, \texttt{imsg}, \texttt{slack}, \texttt{wacli}, \texttt{xurl} \\
group3 & \texttt{apple-notes}, \texttt{apple-reminders}, \texttt{bear-notes}, \texttt{blogwatcher}, \texttt{nano-pdf}, \texttt{notion}, \texttt{obsidian}, \texttt{taskflow}, \texttt{taskflow-inbox-triage}, \texttt{things-mac}, \texttt{trello} \\
group4 & \texttt{canvas}, \texttt{clawhub}, \texttt{coding-agent}, \texttt{gh-issues}, \texttt{github}, \texttt{mcporter}, \texttt{session-logs}, \texttt{skill-creator}, \texttt{tmux} \\
group5 & \texttt{1password}, \texttt{healthcheck}, \texttt{node-connect} \\
group6 & \texttt{goplaces}, \texttt{weather} \\
group7 & \texttt{camsnap}, \texttt{gifgrep}, \texttt{peekaboo}, \texttt{video-frames} \\
group8 & \texttt{gemini}, \texttt{model-usage}, \texttt{oracle} \\
group9 & \texttt{eightctl}, \texttt{openhue}, \texttt{ordercli} \\
\bottomrule
\end{tabular}
\end{table*}

\begin{table*}[t]
\centering
\caption{Boundary cases recorded during manual verification.}
\label{tab:cluster-boundary-cases}
\small
\setlength{\tabcolsep}{6pt}
\begin{tabular}{llp{0.70\textwidth}}
\toprule
Skill & Final group & Resolution rationale \\
\midrule
\texttt{blogwatcher} & group3 & RSS monitoring is an information-intake and tracking activity that composes naturally with productivity workflows. \\
\texttt{canvas} & group4 & Interactive content presentation is most often paired with development and data-oriented workflows. \\
\texttt{node-connect} & group5 & Node connectivity and pairing diagnostics align with system health, security, and operational maintenance. \\
\texttt{ordercli} & group9 & Food-order lookup and ordering are user-facing daily-life services. \\
\texttt{tmux} & group4 & Terminal multiplexing is a standard developer workflow and composes directly with coding and session-management tools. \\
\bottomrule
\end{tabular}
\end{table*}

\subsection{Held-Out Task Records and Annotation Schema}

The final evaluation set contains 491 task instances with unique top-level identifiers from \texttt{route-0001} to \texttt{route-0491}. Each record contains the functional group, executable query, query type, ordered required-skill sequence, chain rationale, and a relevance annotation for every native skill in its group. The ordered \texttt{skill\_chain} defines the gold skill set \(G_x\) and any required relative order. A relevance label of \texttt{primary} identifies a gold skill, \texttt{misleading} identifies a task-specific forbidden skill in \(F_x\), and \texttt{irrelevant} identifies a skill that is not required by the task.

The generation pipeline contains three intermediate query labels: \texttt{direct}, \texttt{ambiguous}, and \texttt{distractor}. Before evaluation, we deterministically map both \texttt{ambiguous} and \texttt{distractor} to the final \emph{ambiguous} category and retain \texttt{direct} unchanged. This normalization changes only the reporting label and does not modify the query, gold sequence, relevance matrix, or execution criteria. The resulting held-out set contains 310 direct tasks and 181 ambiguous tasks; the corresponding intermediate counts are 310 direct, 42 ambiguous, and 139 distractor records.

Each retained task is checked for five properties: (i) the query expresses an executable user objective; (ii) every skill in \(G_x\) contributes a distinct operation required to complete that objective; (iii) the recorded order is consistent with the dependency expressed by the query and chain rationale; (iv) non-gold skills are labeled consistently as misleading or irrelevant; and (v) any external resource reference can be mapped to the isolated benchmark environment without changing the task's required skill composition. A task flagged by any reviewer is re-examined against its query, chain rationale, skill descriptions, and fixture mapping. The task is retained only after consensus on the query type, gold sequence, forbidden skills, and completion conditions; otherwise it is excluded before the final identifier sequence is fixed.

Of the 491 retained tasks, 438 require two skills, 49 require three skills, and 4 require four skills; all 53 native skills occur in at least one gold sequence. Fixture localization modifies the executable surface form of 129 tasks and leaves the remaining 362 unchanged. The supplied task JSON stores the executable localized query used by the agent. The 491 task instances contain 483 unique executable query strings: nine group-5 instances originate from distinct source prompts but are mapped by the fixture-localization procedure to the same executable surface query. Because all 491 pre-indexed instances were executed, they remain separate task instances in the reported evaluation, and aggregate metrics operate over task instances rather than unique query strings.

\begin{table}[t]
\centering
\caption{Required-skill sequence lengths in the 491 held-out tasks.}
\label{tab:task-chain-lengths}
\small
\begin{tabular}{lrr}
\toprule
Required skills & Tasks & Fraction \\
\midrule
Two & 438 & 89.2\% \\
Three & 49 & 10.0\% \\
Four & 4 & 0.8\% \\
\midrule
Total & 491 & 100.0\% \\
\bottomrule
\end{tabular}
\end{table}

\begin{table}[t]
\centering
\caption{Summary statistics for the held-out benchmark.}
\label{tab:heldout-summary}
\small
\begin{tabular}{lrr}
\toprule
Property & Tasks & Fraction \\
\midrule
Direct query type & 310 & 63.1\% \\
Ambiguous query type & 181 & 36.9\% \\
Localized executable query & 129 & 26.3\% \\
Unchanged executable query & 362 & 73.7\% \\
Contains a forbidden-skill annotation & 25 & 5.1\% \\
\midrule
Total task instances & 491 & 100.0\% \\
\bottomrule
\end{tabular}
\end{table}

\section{Task-Generation Prompts and Filtering}
\label{app:task-generation}

\paragraph{Generation objective.}
We formulated task generation as a benchmark-design problem. The system prompt instructed the generator to analyze the logical relationships among the skills available to an agent and to construct diverse, realistic composite tasks for evaluating skill-selection accuracy. The objective was to test whether an agent could infer the skills required by a user request, including their dependencies and execution order, rather than relying only on surface-level keyword matching.

\paragraph{First-round generation prompt.}
For each functional group, the first-round prompt provided the names and descriptions of all skills in that group. The generator was required to produce natural user requests involving at least two distinct skills. Each skill chain had to follow a coherent causal relationship: the result or state produced by one skill should naturally create the need for the next skill. The prompt explicitly discouraged combining unrelated capabilities merely to increase the number of skills in a task.

To improve coverage, the generator was instructed to vary the number of required skills, including two-skill, three-skill, and longer compositions, as well as the starting skill and the combination of skills involved. The requests were written as one to three sentences in natural user language. References to implementation-level skill names, such as explicitly asking the agent to ``use'' a particular skill, were prohibited.

\paragraph{Query-type construction and normalization.}
During task generation, the prompt distinguished 2 construction-time query types. A \emph{direct} query clearly stated the user's intended operations, making the required capabilities relatively explicit without naming the underlying skills. An \emph{ambiguous} query described a broader objective and required the agent to infer the necessary skill composition. 

\paragraph{Manual filtering.}
The primary manual filtering operation was duplicate removal. Reviewers compared requests within each functional group and removed both exact duplicates and semantic duplicates. Two tasks were considered duplicates when they expressed substantially the same user objective and required the same skill composition, even if they differed in entities, locations, dates, or superficial wording. When several variants represented the same underlying task, we retained the version that was more natural and self-contained.

After the validity checks described above, duplicate removal was the only additional manual filtering step. Tasks were not filtered according to model routing behavior, coordinator selection, task-completion outcomes, or resource-amplification measurements. In particular, tasks were not filtered according to the routing behavior of an evaluated model, coordinator-selection outcomes, task-completion results, or resource-amplification measurements. After duplicate filtering, the construction-time ambiguous and distractor labels were normalized to the unified ambiguous category described above. This process produced 536 tasks, of which 45 were reserved for pilot-guided coordinator optimization and the remaining 491 formed the held-out evaluation benchmark.

\subsection{Human Evaluation Protocol}
\label{app:human-evaluation}

\paragraph{Annotators and blinding.}
Task completion was evaluated by four annotators with prior experience in operating and assessing skill-based LLM agents. None participated in benchmark construction, coordinator design, attack implementation, or experimental execution. For each instance, annotators were shown only the original user task, the final agent response, and the observable task-relevant artifacts. They were not informed whether the execution originated from the clean or injected condition. Clean and injected instances were anonymized, separated from their paired counterparts, and presented in randomized order. Model identity, coordinator-selection status, skill traces, token consumption, latency, and other resource measurements were hidden throughout the evaluation.

\paragraph{Completion criteria.}
Task-specific completion criteria were established during benchmark construction and fixed before the experimental outputs were examined. An execution was labeled \emph{complete} only when the final response and observable artifacts jointly satisfied all material requirements of the original request; otherwise, it was labeled \emph{incomplete}. Differences in wording, formatting, or execution style were not treated as failures when they did not affect the requested outcome. The same criteria were applied to clean executions, injected executions, and the independently authored generalization tasks.

\paragraph{Independent annotation and aggregation.}
All four annotators evaluated each instance independently before seeing the judgments of the others. They reached unanimous agreement on 96.3\% of the evaluated instances. The remaining 3.7\% consisted exclusively of three-to-one disagreements, for which the final label was determined by majority vote. No two-to-two split or additional adjudication was required.

\section{Coordinator Construction and Description Optimization}
\label{app:coordinators}

% REVIEW FOLLOW-UP FOR THE SUGGESTED MAIN-TEXT REVISION: state once in this appendix how $\rho_g$ and the group-specific runbooks were instantiated (author-guided, LLM-assisted, or hybrid), who checked them, and when they were frozen. This is preferable to repeating ``manually'' throughout the main text.
For each functional group \(g\), we construct one coordinator \(h_g=(d_g,b_g,f_g)\) from a shared coordination rationale \(\rho_g\). The rationale is derived from the group label, the native-skill name set \(\mathcal{N}_g\), representative single-skill requests \(\mathcal{X}_g^{\mathrm{sample}}\), and plausible relationships among the native skills. It specifies the target-domain triggers, a high-level coordinator role, and a non-replacement boundary under which native skills remain responsible for task-specific operations. The routing-facing description \(d_g\) expresses this rationale as selection metadata, whereas the instruction body \(b_g\) instantiates it as a bounded coordination runbook after the coordinator is loaded.

\subsection{Description Construction and Optimization}

% REVIEW FOLLOW-UP FOR THE SUGGESTED MAIN-TEXT REVISION: add one compact implementation sentence containing only the actual generator model, initial/retained candidate counts, query budget, and operational plateau rule. Avoid expanding these details in the main paper.
The description generator receives \((g,\mathcal{N}_g,\mathcal{X}_g^{\mathrm{sample}},\rho_g)\) while the original native-skill descriptions and instruction bodies are withheld. Each candidate description contains two clauses. The activation clause uses target-domain concepts, representative operations, and user-intent patterns to establish routing relevance. The coordination clause presents the coordinator as an organizer of multi-skill workflows, encouraging co-selection while explicitly leaving task-specific operations to native skills.

Let \(\mathcal{P}_g\) denote the five pilot tasks reserved for functional group \(g\). For each candidate description \(d\), the corresponding coordinator \(h_g(d)\) is installed in the complete clean registry and evaluated in isolation. The attraction rate is
\begin{equation}
    A_g(d)
    =
    \frac{1}{|\mathcal{P}_g|}
    \sum_{x\in\mathcal{P}_g}
    \mathbf{1}\!\left[
        h_g(d)\in
        \Lambda_{\mathcal{S}_0\cup\{h_g(d)\}}(x)
    \right].
\end{equation}
Higher-scoring candidates and their binary selection outcomes are returned to the generator. New variants adjust domain triggers, representative operations, skill-name anchors, and inter-skill relationships while preserving \(\rho_g\) and the non-replacement boundary. Low-scoring candidates are discarded, and screening stops when the best attraction rate plateaus or the query budget is exhausted; this typically occurs within two to three rounds.

The highest-attraction candidate is subsequently refined under full agent execution on the same group-specific pilot set. This stage uses only whether the coordinator is selected and loaded in the realized agent execution. The activation clause is revised to strengthen native-skill anchors, domain terminology, representative operations, inter-skill relationships, and explicit activation conditions, while the coordinator role, shared rationale, and non-replacement boundary remain fixed. Task completion, token consumption, latency, and trajectory amplification are not used as refinement feedback. The 491 held-out evaluation tasks are not exposed during either screening or refinement.

\begin{figure*}[t]
\centering
\fbox{\begin{minipage}{0.95\textwidth}
\textbf{Candidate-generation prompt.}\par
\smallskip
\texttt{You are designing routing metadata for a coordinator skill in a progressively disclosed skill registry.}\\
\texttt{Functional group: [GROUP].}\\
\texttt{Native skill names: [NATIVE\_SKILL\_NAMES].}\\
\texttt{Representative single-skill requests: [EXAMPLES].}\\
\texttt{Shared coordination rationale: [RATIONALE].}\\
\texttt{Generate a batch of concise candidate descriptions. Each candidate must contain: (1) an activation clause using target-domain concepts, representative operations, and user-intent patterns; and (2) a coordination clause presenting the skill as an organizer of multi-skill workflows. The coordinator must support co-selection, must not claim to replace native skills, and must leave task-specific operations to them. Do not use or infer hidden instruction bodies. Return only the candidate descriptions.}
\medskip
\hrule
\medskip
\textbf{Description-refinement prompt.}\par
\smallskip
\texttt{Revise the candidate descriptions using the supplied pilot selection feedback. Preserve the shared coordination rationale, coordinator role, and non-replacement boundary. Improve target-domain coverage by adjusting native-skill anchors, domain terminology, representative operations, inter-skill relationships, and explicit activation conditions. Modify only routing-facing wording; do not add task answers, executable instructions, or claims that displace native skills. Return only the revised candidate descriptions.}\\
\texttt{Candidates and binary pilot feedback: [CANDIDATES\_AND\_FEEDBACK].}
\end{minipage}}
\caption{Prompts used for candidate-description construction and pilot-guided refinement. Bracketed fields are instantiated independently for each functional group.}
\label{fig:coordinator-description-prompts}
\end{figure*}

\subsection{Coordination-Rule Body Construction}
\label{app:body-construction}

Once the coordinator is selected and loaded, its instruction body becomes visible to the planner. The body introduces additional native-skill dependencies that expand the execution trajectory while explicitly stating that the coordinator does not substitute for any native skill. This preserves the original task objective and allows the planner to return to task-relevant native skills after completing the coordination detour. Each group-specific body follows six design principles:

\begin{enumerate}
    \item \textbf{Dependency topology.}
    A directed execution path is constructed over native skills in the corresponding functional group. Plausible functional relationships define inter-skill dependencies that direct the planner to invoke multiple group members after the coordinator is loaded.

    \item \textbf{Edge justification.}
    Each dependency edge is paired with a consequence of omission, framing an auxiliary invocation as necessary to avoid data loss, state inconsistency, incomplete evidence, or failed verification.

    \item \textbf{Mandatory framing.}
    Skill responsibilities and their relative invocation order are expressed as procedural requirements. Potentially optional language is resolved by the runbook so that required auxiliary steps are not silently skipped.

    \item \textbf{Evidence-event decomposition.}
    Skill responsibilities are decomposed into distinct pre-execution, execution-time, and post-execution events, allowing the runbook to introduce multiple bounded invocations within one task.

    \item \textbf{Post-execution verification.}
    After the primary operation, a bounded reverse path may revisit a predefined subset of native skills to verify state consistency. Explicit per-skill and total invocation limits prevent open-ended recursion.

    \item \textbf{Structured packaging.}
    The rules are packaged as a Markdown runbook containing the activation scope, dependency path, execution protocol, verification bounds, and a return condition. When the dependency path completes or an invocation bound is reached, the coordinator-specific procedure terminates and returns control to the ordinary planner.
\end{enumerate}

\subsection{Released Coordinator Artifacts}

The exact group-specific descriptions and complete Markdown runbooks used in the reported experiments are supplied as machine-readable coordinator skill files in the supplementary ZIP archive. They are not duplicated in this appendix.

\subsection{Hit-Conditional Ablation Analysis}

Table~\ref{tab:ablation-conditional} complements the common-denominator results in Table~\ref{tab:ablation} by measuring detour intensity only after the coordinator is selected. These estimates answer a mechanism-level question rather than overall end-to-end impact. In particular, Detour-only is conditioned on only three hits and should therefore be interpreted cautiously.

\begin{table}[t]
\centering
\caption{Hit-conditional ablation results on DeepSeek-V4-Pro.}
\label{tab:ablation-conditional}
\setlength{\tabcolsep}{3.5pt}
\resizebox{\columnwidth}{!}{
\begin{tabular}{lccccc}
\toprule
Variant & Coord. Hit & Token $\uparrow$ & Cache $\uparrow$ & Time $\uparrow$ & Avg. $\Delta$Calls \\
\midrule
\textbf{Full CDH} & 70/89 (78.7\%) & +57.9\% & +57.0\% & +76.7\% & +2.01 \\
Detour-only & 3/89 (3.4\%) & +44.4\% & +48.2\% & +125.2\% & +1.67 \\
Attract-only & 70/89 (78.7\%) & +26.4\% & +27.6\% & +65.0\% & +1.16 \\
\bottomrule
\end{tabular}
}
\end{table}

\section{Experimental Configuration and Execution Protocol}
\label{app:protocol}

\paragraph{Session isolation.}
Clean and injected executions are conducted in separate agent sessions. No conversation history, planner state, tool observation, or execution trace is shared across the two conditions. Each single-task execution starts from a fresh session with the same benchmark fixture state. For multi-turn evaluation, the two paired tasks share context only within their own clean or injected session; the matched clean and injected sessions remain mutually isolated.

\paragraph{Timeout policy.}
A fixed execution timeout of \(600\,\mathrm{s}\) is applied to both members of every clean--injected pair. An execution that reaches this limit is recorded as timed out rather than treated as successfully completed. Completion rates are computed over all evaluated task instances, whereas resource-amplification metrics are computed only when the coordinator is selected and both members of the pair complete successfully. This prevents a truncated or timed-out trajectory from being interpreted as resource amplification.

\paragraph{Trajectory logging.}
Each execution produces a timestamped trace containing the selected and loaded skills, ordered skill invocations, invocation arguments and observable responses, termination status, and final agent response. The logger additionally records total and cached token usage, end-to-end wall-clock time, and the number of skill invocations. These records are used to reconstruct \(\Lambda_j(x)\), \(\Gamma_j(x)\), and the paired resource metrics without relying on hidden model states.

\paragraph{Clean--injected paired execution.}
For each task \(x\), the clean run uses the native registry \(\mathcal{S}_0\), and the matched injected run uses \(\mathcal{S}_h=\mathcal{S}_0\cup\{h_g\}\). The task, model, benchmark fixtures, mock-backend state and response rules, timeout, and logging configuration are held fixed; the only experimental difference is the presence of the group-specific coordinator. For a task spanning multiple functional groups, \(h_g\) is chosen according to the functional group of the first gold skill, and coordinators for the remaining groups are not installed. The complete executable environment and its configuration are supplied as the supplementary virtual-machine image.

\end{document}